\documentstyle[aps]{revtex}
\newcommand{\beq}{\begin{equation}}
\newcommand{\eeq}{\end{equation}}
\newcommand{\bi}{\begin{itemize}}
\newcommand{\ei}{\end{itemize}}
\newcommand{\beqar}{\begin{eqnarray}}
\newcommand{\eeqar}{\end{eqnarray}}

\newcommand{\al}{\alpha}

\newcommand{\gv}{\mathrm{ \ GeV \ }}
\newcommand{\nn}{\nonumber}

\newcommand{\crn}{\nn \\}
\newcommand{\wz}{\sqrt{2}}
\newcommand{\bea}{\begin{eqnarray}}
\newcommand{\eea}{\end{eqnarray}}
\newcommand{\mw}{M_W^2}
\newcommand{\mz}{M_Z^2}
\newcommand{\dal}{\Delta \alpha}
\newcommand{\der}{\Delta r}

\newcommand{\sing}{\sin^2 \Theta_W}

\newcommand{\sinf}{\sin^2 \Theta^f_{\mathrm{eff}}}
\newcommand{\cose}{\cos^2 \Theta^\ell_{\mathrm{eff}}}
\newcommand{\sine}{\sin^2 \Theta^\ell_{\mathrm{eff}}}

\newcommand{\sinb}{\sin^2 \Theta^b_{\mathrm{eff}}}
\newcommand{\sinW}{\sin^2 \Theta_W}
\newcommand{\sinWz}{\sin^2 \Theta_W^0}

\newcommand{\ctanWz}{\frac{\cos^2 \Theta_W^0}{\sin^2 \Theta_W^0}}
\newcommand{\droirr}{(\Delta \rho )_{\rm irr}}
\newcommand{\dro}{\Delta\rho}
\newcommand{\Gmu}{G_{\mu}}
\newcommand{\ra}{\rightarrow}

\let\epsilon\varepsilon

\begin{document}

\title{Confronting electroweak precision measurements with New Physics
models}

\author{M. Czakon$^a$, J. Gluza$^{a,b}$, F. Jegerlehner$^b$ and M. Zra\l ek$^a$}

\address{$^a$ Department of Field Theory and Particle Physics, Institute 
of Physics, University of
Silesia, \\ Uniwersytecka 4, PL-40-007 Katowice, Poland \\
$^{b}$  DESY Zeuthen, Platanenallee 6, D-15738 Zeuthen, Germany}

\maketitle

\begin{abstract}\noindent

Precision experiments, such as those performed at LEP and SLC,
offer us an excellent opportunity to constrain extended gauge model
parameters. To this end, it is often assumed, that in order to obtain
more reliable estimates, one should include the sizable one--loop
Standard Model (SM) corrections, which modify the $Z^0$ couplings as
well as other observables. This conviction is based on the belief that
the higher order contributions from the ``extension sector'' will be
numerically small. However, the structure of
higher order corrections can be quite different when comparing the SM with
its extension, thus one should avoid assumptions which do not care
about such facts. This is the case for all models with 
$\rho_{\rm tree} \equiv M_W^2/(M_{Z}^2\cos^2{\Theta_W}) \neq 1$.
As an example, both the manifest
left-right symmetric model and the $SU(2)_L \otimes
U(1)_Y \otimes \tilde{U}(1)$ model, with an additional
$Z'$ boson, are discussed and special attention  to the top 
contribution to $\Delta \rho$ is given. We conclude  
that the only sensible way to confront a model with the
experimental data is to renormalize it self-consistently, if
not, parameters which depend strongly on quantum effects should be
left free in fits, though essential physics is lost in this way. 
We should note that arguments given here allow us 
to state that at the level of loop corrections (indirect effects)
there is nothing like a ``model independent global analysis'' of the data.

\end{abstract}

\section{Introduction}
 
It is a remarkable fact, that precise theoretical predictions of the
electroweak SM, obtained after taking into account one-, two-, or even
in some cases three- loop effects, fully agree with all experimental
data which have been accumulated so far and which have reached a
surprisingly high level of precision~\cite{hol}.  Moreover, these
theoretical calculations have a high indirect predictive power because
of the substantial sensitivity to non--decoupling heavy particle
effects. A potentially large top quark contribution to boson
self-energies has been recognized long time ago~\cite{vel}.  Based on
this, the top mass has been estimated quite accurately
($m_t^{\rm ind}=170(184)\pm7 \gv$, assuming $M_H=M_Z(300 \gv)$)~\cite{new}
prior to its direct determination ($m_t^{\rm dir}=173.8\pm5.2 \gv$)
which confirmed the indirect result not so long ago~\cite{top}.  Now,
with the top quark at hand, the only not yet discovered particle which
is required in the SM, the Higgs boson, can be studied. At present,
the indirect bound after inclusion of the relevant higher order
corrections to the $Z^0$--peak observables implies $m_H< 262$ GeV at
95\% C.L.~\cite{deg}.

It could be that better and better agreement between SM theory and
experiments will follow the increasing sophistication of perturbative
calculations. In the framework of the SM, this is a logic and obvious
possibility.

In the following, let us focus on a different scenario.  There are
many arguments against the SM to herald in the ultimate theory of
elementary particles.  We believe that, beyond the SM regime, at
higher energies, new physics will show up. Precision experiments
provide us an important tool to find its remnants already at todays
energies. They have been analyzed in the context of many different
models, e.g, those which include an additional $Z'$ boson. For details
we refer to~\cite{lan}.  It is customary to assume that extended
models can be constrained in particular by the neutral current (NC)
data, through their modified tree level $Z^0$ couplings and improved
by radiative corrections from the SM.  Contributions from the
heavy non-standard sector  seem to be negligible in
a first approximation.  However, the situation in general is more
complicated and this ``standard'' approach can be misleading. Before going
further we should  make this point clearer. In GUT models, typically, per
construction a gauge hierarchy exists 
\cite{gild}: a Higgs field exhibites a small vacuum expectation value (VEV) $v$
determines the SM particle mass spectrum and another Higgs field with a 
large VEV $V$ 
generates the super heavy sector. Decoupling theory states \cite{kaz} that once a
proper identification of the light and of the heavy particles at tree level is
done then such a division will be maintained in any order of perturbative
calculations (all the super heavy particle effects enter at most  as
logarithmic corrections to the light particle  effects). However, in
phenomenological applications we have no direct experimental access to the
parameters of the heavy sector $(V,M_{H_i},...)$ but only to some effective low
energy parameters, like for instance $\rho$ parameter
which is also a function of the parameters of the heavy sector. If we constrain
the low energy effective parameter by experiment (in some physical on-shell
renormalization scheme) then we in general set up boundary conditions which are
not compatible with the set up of a gauge hierarchy and the just mentioned
decoupling theory does not work. This has further consequences.
After letting the superheavy  masses go to infinity,
 the low energy effective
theory (assuming light fields are the same as in the SM) is not any longer
renormalizable, 
much in the same way as the low energy
effective four fermion interactions are nonrenormalizable if we fix $G_{\mu}$
and let $M_W \to \infty$ (which requires $g \to \infty$ simultaneously).

\section{Discussion}

To outline our point of view let us consider left (L) -- right (R)
symmetric models (LRM) with gauge group $SU(2)_L \otimes SU(2)_R \otimes
U(1)_{B-L}$ which are manifestly LR--symmetric before the symmetry is
broken by the appropriate Higgs mechanism~\cite{man}. These models
have all the necessary features of a large class of extended models,
and some results at the one-loop level have lately been
obtained~\cite{pil,cza} which are applicable to LEP/SLC physics.

Let us start by considering the $Z^0$ partial decay widths and
forward-backward asymmetries,
theoretically described by the following relations~\cite{lep}:

\bea
\label{width}
\Gamma_{f\bar{f}}&=&\frac{N_c^fG_F M_Z^3 }{6 \pi
\sqrt{2}} \, \beta \left(
\frac{3-\beta^2}{2} v_f^2+ \beta^2 a_f^2 \right) K_{QCD} K_{QED},\\
A_{\mathrm{FB}}^f&=&\frac34 A_e A_f\;,\;\;A_f=\frac{2v_f a_f}
{\left(v_f^2+a_f^2\right)}
\label{FBasym}
\eea
where $N_c^f$ is the color factor, $\beta$ the fermion velocity, the
$K$ factors take into account electromagnetic and strong corrections,
and $v_f$ and $a_f$ are
vector and axial fermion couplings.  In the LRM model these can be
written in the simple and compact form ($T^3_f,Q_f$ being fermion's
isospin and charge, respectively):
\begin{eqnarray}
v_f&=& \sqrt{\rho_{\rm eff}^f}\,(T^3_f-2 Q_f \sinf)
(\cos{\phi}-\sin{\phi}/\sqrt{\cos{2 \Theta_W}}\:) 
\label{vec}
\\ a_f&=&\sqrt{\rho_{\rm eff}^f}\,T^3_f\,
(\cos{\phi}+\sin{\phi}\,\sqrt{\cos{2 \Theta_W}}\:)\;.
\label{axi}
\end{eqnarray}
Here $\phi$ is the $Z^0-Z'$ mixing angle and the two other angles are
connected to the effective weak mixing parameter $\sinf$ in the NC at the
$Z^0$ resonance (for which (\ref{vec}) is the defining equation) and
the weak mixing angle $\Theta_W$ defined via the vector boson masses by
\bea
\sinW =1-\frac{M_W^2}{\rho_0 M_Z^2}\;. 
\label{tree}
\eea
While the $\rho$-parameter is unity at the tree level in the SM, it
differs from unity in many extended models: $\rho_{\rm tree} \equiv \rho_0
\neq 1$. Let us assume that higher order effects are really small and can be
gathered by SM like relations
\bea
\rho&=&\frac{\rho_0}{(1-\Delta \rho)} \nn \\
\rho_{\rm eff}^f&=&\rho\,(1+\Delta
\rho_{\rm rem}^f)\;.
\label{rho}
\eea
In the LR model $\rho_0$ should be understood as $\rho_0/\rho^{\pm}$ where
$\rho_0$ is given by $\rho_0 =1+\sin^2 \phi \left(
M^2_{Z_2}/M^2_{Z_1}-1 \right) $ and is due to the $Z$--$Z'$ mixing and
$\rho^{\pm}=1+ \sin^2 \phi_{\pm}\:(M^2_{W_2}/M^2_{W_1}-1)$ is due to the
$W$--$W'$ mixing.

In terms of the input parameters $\al$, $G_F$, $M_Z$,... with 
 $A_Z=\frac{\pi \alpha(M_Z)}{\sqrt{2} G_F}$ and $\alpha(M_Z)=
\frac{\al}{1-\Delta \alpha}$ we can predict

\begin{eqnarray}
\sin^2{\Theta_W}&=& 
\frac{1}{2} \left[ 1-\sqrt{1-\frac{4 A_Z}{\rho  M_Z^2}
\left(1+\Delta r_{\rm rem}\right)} \; \right] 
\label{sinW}
\\ \nonumber
\\ 
\sinf &=&
\frac{1}{2} \left[ 1-\sqrt{1-\frac{4 A_Z}{\rho_{\rm eff}^f M_Z^2}
\,\left(1 + \Delta r^f_{\rm rem}
\right)}\; \right]\;, 
\label{sinf}
\end{eqnarray}
with leading higher order corrections incorporated in resummed
form~\cite{CHJ}.  Let us put $\phi=0$, so that pure SM physics is
restored. Then the terms $\Delta \alpha$, $\Delta \rho$, $\Delta
\rho_{\rm rem}^f$, $\Delta r_{\rm rem}$, $\Delta r^f_{\rm rem}$
include SM radiative corrections to the $Z^0$ and muon
physics~\cite{lep}. These depend on many details, for instance, the
$f$ superscript means that actually $\sinf$ and $\rho_{\rm eff}^f$ are
not universal quantities but differs for each fermion flavor produced
at the $Z^0$ resonance through flavor specific vertex (and box)
effects.  The flavor dependence, however, is relatively small except
for $f=b$ which requires separate treatment. Appealing to lepton
universality, we denote the leptonic weak mixing parameter by $\sine$
($\ell=e,\mu$ or $\tau$). Some of the radiative corrections are
dominant.  For instance, in Eq.~(\ref{sinW}) the two leading effects
have been incorporated by including the running of the fine structure
constant (shift by $\Delta \alpha$) from low to high ($Z$-mass)
energies and the renormalization of $\rho_0=1$ by the large mass
splitting between top and bottom quarks in boson self-energies (shift
by $\Delta \rho$):
\begin{equation}
 \Delta \rho = \Delta \rho^{\rm top}+\Delta \rho_{\rm rem}\;,\;\; 
\Delta \rho^{\rm top} = 3 x_t\;,\;\;
x_t\equiv \frac{\sqrt{2}G_F}{16 \pi^2} m_t^2 
\label{dtop}
\end{equation}
For $f \neq b$, all other contributions indexed by ``rem'' are smaller
remainder terms, e.g., $\Delta r^f_{\rm rem}$ is the remainder
gathering non-leading effects from boson self-energies, vertices and
boxes. In the case $f=b$ there is a leading top mass correction coming
from the $Zb\bar{b}$ vertex~\cite{ABR} which can be incorporated as
\bea
\rho_{\rm eff}^b=\rho_{\rm eff}^\ell \,(1+\tau_b)^2 \crn
\sinb = \sine / (1+\tau_b) 
\label{zbb}
\eea
with $\tau_b=-2 x_t$ (see (\ref{dtop})).
All correction factors influence $\Gamma_{f \bar{f}}$,
$A_{\mathrm{FB}}^f$ given in Eqs.~(\ref{width},\ref{FBasym}), as well
as other observables.

Now, let us switch on ``new physics'' again $(\phi \neq 0)$.  The
question is (apart from coupling modifications) what is going to
change in the loop effects.  As written in the introduction, the
``canonical'' answer is~\cite{afe} (here we refer only to papers where
LRM have been considered): Eqs.~(\ref{rho}-\ref{dtop}) will not be
changed, except for negligible contributions affecting the sub-leading
terms. The leading behavior will be governed by the SM.

However, beyond the tree level, as shown in~\cite{cza}, a substantial
part of the relevant radiative corrections change completely and there
is only a weak relationship between the radiative corrections of the SM
and the new physics model (NPM=extended SM). While corrections like
$\Delta \alpha$ are universal others may change dramatically, in
particular the non--decoupling heavy particle effects. For instance,
one of the most important one-loop terms, $\Delta \rho^{\rm top}$ looses
its $m_t^2$ dependence, namely, in the LR model we obtain
\begin{equation}
 \Delta \rho^{\rm top}_{\rm LR} =
\frac{\sqrt{2}G_F}{8\pi^2} c_W^2\left( \frac{c_W^2}{s_W^
2}-1 \right)\frac{M_{W_1}^2}
{M_{W_2}^2-M_{W_1}^2}3m_t^2
\label{rhoLR}
\end{equation}
as a leading term. For a $W_2$ boson mass of the order of $400$ GeV
or larger this contribution is much smaller than the SM one, 
actually even smaller than the SM logarithmic terms. Besides this,
other particles like heavy neutrinos and heavy scalars~\cite{pil,cza}
influence substantially the sub-leading terms in Eqs.~(\ref{rho}-\ref{dtop}).

The traditional philosophy simply breaks down.  When fitting
parameters within the framework of a NPM, e.g. the $Z^0-Z'$ mixing
parameter $\phi$, the only way of including one-loop effects is to
renormalize the whole model. Except from universal corrections like
the QED shift $\dal$, it is not legitimate to use radiative
corrections from the SM for its extension unless $\rho_0$ remains
unity. Affected are in particular the zero momentum gauge boson
contributions. Although at low energies and at tree level the LRM seems
to be effectively equivalent to the SM ($\phi,\phi_{\pm} \to 0$ and
$M_{Z_2},M_{W_2} \to \infty$), radiative corrections can be quite
different and do not follow this naive expectation 
(see Eq.~(\ref{rhoLR}) and
$M_{W_2} \to \infty$)\footnote{As discussed at the end of the Introduction
we should be careful in refering to decoupling in the limit
 $M_{W_2} \to \infty$. In reality we
 fix $\Delta \rho^{\rm top}_{\rm LR}$ to experimental data, which means
also that a limit  $M_{W_2} \to \infty $ not necessarily is allowed any longer.}.

The crucial point is that associated
with the additional free parameters there are new divergences and
hence new subtractions needed.  Then Eqs.~(\ref{rho}-\ref{dtop}) will
get additional contributions and now will be functions of the extended
set of input parameters (SM parameters plus
$\phi,M_{W_2},M_{Z_2},...$). 
Let us note, that the 
naively written 
one-loop level definition of  $\sinf$ in Eq.~(\ref{sinf})
should also be different from  its SM  structure.
The LRM angle $\phi$ can be fixed at tree level by:
\begin{equation}
\sin{2\phi}=-\frac{g^2 \sqrt{\cos{2\Theta_W}} \left[ \left( g^2+g'^2 \right)
\left( M_{W_2}^2+M_{W_1}^2 \right) -\frac{1}{2}g^2 \left( M_{Z_1}^2+M_{Z_2}^2
\right) \right] }{ \cos^2{\Theta_W} \left( M_{Z_2}^2-M_{Z_1}^2
\right) g^2  \left( \frac{1}{2}g^2+g'^2 \right)}
\label{philr}
\end{equation}
and  extraction of $\sinf$ from $Z$-fermion
couplings Eq.~(\ref{vec}) at the one-loop level
will also include its renormalization.
The same touches the $\sin^2{\Theta_W}$ definition 
Eqs.~(\ref{tree},\ref{sinW}) ,
where $\rho_{\rm tree} \neq 1$ is present (see \cite{cza} for the renormalization of the 
$\sin^2{\Theta_W}$ parameter).

The observation that the structure of higher order effects is highly
model dependent was pointed out long time ago in~\cite{LN} for the
case of models with an enhanced Higgs sector (the so called
``unconstrained'' extended models) for which the custodial symmetry
exhibited by the SM Higgs is violated at the tree level, causing
$\rho_{\rm tree} \neq 1$. In~\cite{jeg} it was shown in general, how
the SM radiative corrections are modified in models which require a
direct or indirect renormalization of the $\rho$--parameter. See
\cite{blank} for an analysis of precision observables in a SM enhanced
by an additional Higgs triplet.  In any case, if $\rho$ is itself a
free parameter or a function of other input parameters, the quadratic
top mass contributions coming from self-energy diagrams are lost by
the required subtraction and only logarithmic top mass dependences
remain.  The dependence on the Higgs mass is also affected
substantially (see the Appendix for details). Hence, in models with
$\rho_{\rm tree} \neq 1$ the LEP/SLC indirect top mass limits become
obsolete. Such models are unable to explain why the direct top mass
agrees with the one obtained from precision measurements of the loop
effects in $\Delta \rho$. The coincidence $m_t^{\rm ind} \simeq
m_t^{\rm dir}$ obtained by SM fits has a meaning only when $\rho$ is a
finite calculable quantity, which requires $\rho_{\rm tree} = 1$, like
in the SM or in its minimal supersymmetric extension. In the case of
the LRM, which we have discussed before, the phenomenon of a complete
change in the large $m_t$ behavior to~(\ref{rhoLR}) was obtained in a
different renormalization scheme which did not treat $\rho_{\rm tree}$
itself as an independent parameter. In contrast to the $m_t^2$
dependence originating in the $W$ and $Z$ self-energies at zero
momentum, the $m_t^2$ dependence of the $Zb\bar{b}$ vertex is not (or
little) affected when going to an extended model. Therefore, the
observables including $b$ quark contributions, like
$\Gamma_{b\bar{b}}$, $A_{\mathrm{FB}}^b$, the $Z$ width or the $Z$
peak cross-section, still exhibit strong $m_t$ dependencies (now very
different from the ones in the SM) which allow to get good indirect
$m_t$ bounds~\cite{blank,lang}. However, there is no good reason why
the new bounds should coincide with the ones obtained in the SM. This
does not necessarily mean that one cannot obtain equally good global
fits, because in the extended model more free parameters are at our
disposal ($\rho_0$ free fits~\cite{lang}).

The mentioned ``instability of quantum effects'' may also be observed
in rather simple modifications of the SM, like, the $SU(2)_L \otimes
U(1)_Y \otimes \tilde{U}(1)$ models, which often arise as the low
energy limit of interesting GUT's, and which exhibit an additional
$Z'$ boson mixing with the $Z^0$. We may restrict ourselves to
consider the \textit{constrained} version, where Higgs bosons transform as
doublets or singlets of $SU(2)_L$. Aspects of the renormalization of
such models have been considered in Ref.~\cite{DS}.
 
If $Z'$ mixes with $Z^0$ then we obtain neutral vector bosons of
masses $M_{Z_1}(\leq M_{Z^0})$ and $M_{Z_2}(\geq M_{Z^0})$ and at the tree 
level the $Z_1-Z_2$ mixing angle $\phi$ is fixed by:
\begin{equation}
\tan^2{\phi}=\frac{M_W^2/\cos^2{\Theta_W}-M_{Z_1}^2}{M_{Z_2}^2-M_W^2/
\cos^2{\Theta_W}}
\end{equation}
or, equivalently:
\begin{equation}
\rho_{\rm tree} \equiv \rho_0 \equiv \frac{M_W^2}{M_{Z_1}^2\cos^2{\Theta_W}} =
 1+\sin^2{\phi} \left( \frac{M_{Z_2}^2}{M_{Z_1}^2}-1 \right)> 1\;.
\end{equation}

In~\cite{DS} $\sin^2{\Theta_W}$ has been calculated in terms of
$\alpha, G_F$ and $M_W$ at one-loop order
\begin{equation}
\sin^2{\Theta_W}=\frac{\pi \alpha}{\sqrt{2} G_F M_W^2} (1 + \Delta \tilde{r})
\end{equation}
with the conclusion that $\Delta \tilde{r} \simeq \Delta r^{\rm SM}$
up to negligible corrections, in a scheme where continuity in the
limit $\phi \rightarrow 0$ is imposed by hand. Note that this
relation, which derives from the charged current (CC) muon decay, is
not modified at the tree level.  Thus $\sinW \simeq \sinW ^{\rm SM}$
when calculated in terms of $\al$, $\Gmu$, $M_W$ and the subtraction
is imposed at $\phi=0$.

However, if we calculate $\sin^2{\Theta_W}$ in terms of
$\alpha, G_F$ and $M_Z$ (the standard input parameters for precision
calculations), again at one-loop order, we have
\begin{equation}
\sin^2{\Theta_W} \cos^2{\Theta_W} = \frac{\pi \alpha}{\sqrt{2}
G_F\rho_0 
M_Z^2} (1 + \Delta \tilde{r})
\end{equation}
which is modified by the appearance of the new parameter $\rho_0$,
which has to be renormalized now as well. Since $\rho_0$ acts as a
free parameter we cannot get any longer the $m_t$ bounds of the SM. In
the commonly accepted procedure one would argue as follows: in linear
approximation, due to $\rho_0 =1+\dro_0 \neq 1$ we get effectively an
extra classical contribution 
\bea \delta \der = -\ctanWz \dro_0, \;\;
\dro_0 = \sin^2 \phi \left( \frac{M^2_{Z_2}}{M^2_{Z_1}}-1 \right)
\eea 
where
\bea
\sinWz=1-\frac{M_W^2}{M_Z^2}\;.
\eea
Thus it looks as if we would substitute in (\ref{rho}) 
\bea
\dro^{\rm top} \ra \dro^{\rm top} + \dro_0 
\eea 
with both contributions positive. Formally, one seems to be able to
constrain both $m_t$ and $\rho_0$. After a full one-loop
renormalization of the NPM a term $\dro^{\rm top} \sim m_t^2$ is
absent, however, and the conventional recipe breaks down (see the
Appendix for details).

We conclude that self-consistent constraints on the NPM parameters can be 
obtained only by a consequent order by order analysis of the model.

The question which remains is the following: can we make reasonable
fits of the new parameters without any knowledge of the radiative
corrections in the NPM?  The answer is positive.

Let us take the LEP/SLC data~\cite{new}

\begin{eqnarray*}
M_Z&=&91.1867 \pm 0.0021\; \mbox{\rm GeV} \\
\Gamma_Z&=& 2.4939 \pm 0.0024\; \mbox{\rm GeV} \\
\sigma_h^0&=& 41.491 \pm 0.058\; \mbox{\rm nb} \\
R_\ell&=&20.765 \pm 0.026 \\
A_{FB}^{0,\ell}&=&0.01683 \pm 0.00096
\end{eqnarray*}

They have  been extracted from the line-shape and lepton asymmetries.
We will also use
$A_\ell$, $R_b^0$, $R_c^0$, $A_{FB}^{0,b}$,
$A_{FB}^{0,c}$, $A_b$, $A_c$ (values, correlation matrices and definitions
are to be found in~\cite{new}). 
The  important point is that all of them are 
expressible through Eqs.~(\ref{width}-\ref{axi}).

According to our approach $\sin{\Theta_W}$, $\sinf$ should be left as
free parameters.  But a closer look at Eqs.~(\ref{vec},\ref{axi}),
leads to the conclusion that their values can not be separated from
$\phi$ (the $\chi^2$ minimization procedure~\cite{min} would break
down).  This is why we have to tune rough starting values for the
weak mixing parameters.  Instead of~\cite{new}:
\begin{eqnarray}
\sine (Q_{FB})&=& 0.2321 \pm 0.001 \\
\sin^2{\Theta_W}&=& 0.2254  \pm 0.0021
\end{eqnarray}
which are  extracted from experiment,
the following starting points to the $\chi^2$ minimization procedure
 
\begin{eqnarray}
\sine&=& 0.230 \pm 0.01 \label{st} \\
\sin^2{\Theta_W}&=& 0.225  \pm 0.01
\end{eqnarray}
are taken.

$\rho_{\rm eff}^f$ is also to be taken as a free parameter.
Technically, we know that results connected to the heavy b quark at
LEP differ from those of other fermions. We take it into account and
introduce two more free parameters, namely $\rho_{\rm eff}^{b}$ and
$\sinb$ (with the starting value as given in
Eq.~(\ref{st})).

To sum up, we have 18 physical data ($M_Z$, $\Gamma_Z$, $\sigma_h^0$,
$R_\ell$, $A_{FB}^{0,\ell}$, $A_\ell$, $R_b^0$, $R_c^0$, $A_{FB}^{0,b}$,
$A_{FB}^{0,c}$, $A_b$, $A_c$, $\sine$, $\sinb$, $\sin^2{\Theta_W}$,
$m_t$, $\alpha_{s}$, $M_W$) as a function of 3 completely free
parameters $\rho_{\rm eff}^\ell$, $\rho_{\rm eff}^{b}$,$\phi$.

The $\chi^2$ minimization procedure
gives (at 90 \% C.L.):
\begin{eqnarray}
| \phi | & \leq & 0.003 \label{con} \\
\rho_{\rm eff}^\ell & = & 1.005 \pm 0.004 \label{rhot1} \\
\rho_{\rm eff}^{b} & =& 1.002 \pm 0.028 \label{rhot2} 
\end{eqnarray}
and
\begin{eqnarray}
\sine & =& 0.232 \pm 0.001 \label{sine} \\
\sinb & =& 0.236 \pm 0.021  \label{sinb} \\
\sin^2{\Theta_W} &=& 0.2254 \pm 0.0045 \label{tw}
\end{eqnarray}

If we assume, as already discussed,  
that the $Zb\bar{b}$ vertex is not affected too much in NPM then
the relations given in Eq.~(\ref{zbb}) hold and an upper 
limit on the top mass can be derived.
We get within given errors from Eqs.~(\ref{rhot1},\ref{rhot2}):
\begin{equation}
\frac{\rho_{\rm eff}^b}{\rho_{\rm eff}^e} = (1+\tau_b)^2 \geq 0.965
\end{equation} 
from which $m_t \leq 290$ GeV follows (a weaker limit comes from the 
$\sine / \sinb $ ratio, Eqs.~(\ref{sine},\ref{sinb})). 
See also the discussion in Ref.~\cite{lang}.

In the frame of the LR model, for $M_{Z_2}>>M_{Z_1}$ we may use the 
approximate relation~\cite{plb}
\begin{equation}
\phi  \simeq \sqrt{2 \cos{\Theta_W}} \frac{M_{Z_1}^2}{M_{Z_2}^2}
\end{equation}
in order to obtain the $Z_2$ mass bound
\begin{equation}
M_{Z_2} \geq 1420\; \mbox{\rm GeV}\;.
\end{equation}  

This is a quite strong constraint 
\footnote{Our analysis should be taken as an illustration only.
Our approach is not fully self-consistent. Some of the $Z^0$
parameters used here are so-called pseudo-observables, which have been
extracted from experimental data utilizing SM radiative corrections
\cite{bar}.  We could in principle extract the $Z^0$ parameters from
experimental data using the ZFITTER program \cite{tor} leaving,
according to our approach, model-dependent radiative corrections as
free numbers and see precisely the difference in fits.  Also
$\gamma-Z$ interference should be taken into account in the
appropriate manner.}  (see~\cite{erl} for a comprehensive analysis
including also the low energy data).
However, we should stress here that treating $\sin^2{\Theta_W}$ and
$\sin^2{\Theta_{\rm eff}^f}$ as ``black boxes'' we lost essential
physical information on the NPM. In reality, at loop level,
$\sin^2{\Theta_W}$ and $\sin^2{\Theta_{\rm eff}^f}$ are complicated
functions of new parameters e.g. $\phi,\;\;M_{Z_2},\;\;M_{W_2}$, heavy
neutrinos, extra Higgs particles.  We
do not know what is the relation between the result obtained in
Eqs.~(\ref{con})-(\ref{tw}) and those which would come from the full one-loop
analysis.

\section{Conclusions}

To summarize, fitting precision data requires precise predictions
(including the relevant higher order effects) to be confronted with
data, i.e., for conclusive comparisons the precision of data and
theory have to match as far as possible.  For example, fitting the
electroweak data with SM tree level predictions only, would rule out
the SM, while including radiative corrections leads to perfect
agreement. These rules apply as well for any extension of the SM. Such
NPM exhibit additional free parameters, so that parameters of the SM,
which may be substantially shifted by higher order SM corrections,
turn into free parameters in the NPM. It is thus obvious that taking
into account just the SM radiative corrections plus the tree level
extension cannot make sense, in general. This is the case in
particular for all $\rho_{\rm tree} \neq 1$ extensions.  In our
opinion, there is much more model dependences of global fits and their
interpretation than usually presumed. As an example, the $S$, $T$, $U$
parameter description of physics beyound  the SM \cite{stu,AlBa} 
directly only applies to
$\rho=1$ extensions, like models with additional fermion families
(already discussed in~\cite{vel}), additional scalar singlets and
doublets, massive neutrinos which might exhibit $\nu$-mixing and
supersymmetric extensions of the SM. For $\rho\neq1$ extensions our
discussion concerning $\Delta \rho$ and the $m_t$ bounds applies
directly to $T$ which is defined as $\Delta \rho /\al$.  $S$ and $U$
are scale sensitive quantities which are \textit{expected} to survive
modifications in the renormalization procedure. The problem here is
that the gauge boson self-energies which are intended to be described
by these parameters are not observables themselves. They cannot be
separated in general from vertex and box corrections. See also the
discussion within the effective Lagrangian approach~\cite{nyf} for
this point. One of the most important results of the electroweak
precision measurements is the fact that $\rho$ is very close to its SM
prediction. All models with $\rho_0 \neq 1$ have a severe fine tuning
problem: why does the value of the ``$\rho_0$ free'' fits yield a
result which by accident is very close to the SM prediction?

\begin{center}
{\large \bf Acknowledgments}
\end{center}
We would like to thank A. Nyffeler for helpful discussion.
This work was supported by the Polish Committee for Scientific Research under 
Grants Nos.~2P03B08414 and 2P03B04215. 
J.G. would like to thank the Alexander von Humboldt-Stiftung for a 
fellowship.

\begin{center}
{\large \bf 
Appendix: Modification of the SM  top
quark and Higgs boson contributions in extensions of the SM with $\rho \neq 1$}
\end{center}

One of the crucial features of the SM is the validity of the
relationship \bea \rho=\frac{G_{NC}}{G_{CC}}=\frac{\mw}{\mz
\cos^2{\Theta_W^0}}=1\;,\;\; \cos^2{\Theta_W^0}=\frac{g^2}{g^2+g^{'2}} \eea at the tree
level. As discussed in the main text, many extensions of the minimal SM share this property with the
SM. 
For all
these models \bea \frac{G_{NC}}{G_{CC}}(0)=\rho=\frac{1}{1-\dro}
\label{deltarho}
\eea
is a calculable quantity which is sensitive to weak hypercharge breaking and
weak isospin breaking  due to mass splittings of multiplets.
Here we mention that if $\rho_0=\rho_{\rm tree} \neq 1$ one
should consequently replace 
\bea
\sin^2{\Theta_W^0} &\rightarrow & \sing=(e/g)^2=1-\frac{M_W^2}{\rho_0 M_Z^2} \\
\der &\rightarrow & \der_g =1-\frac{\pi \al}{\sqrt{2}\Gmu M_W^2}\frac{1}
{\sing} \nn
\eea
in all SM formulae. If we define $\dro_0$
in analogy to Eq. (\ref{deltarho}) by
\bea
\rho_0=\frac{1}{1-\dro_0} \nn
\eea
we have
\bea
\sing=\sinWz \; \left( 1+\ctanWz \dro_0 \right) \nn
\eea
and hence the {\em exact} relation
\bea
\frac{1}{1-\der}=\frac{1}{1-\der_g}\; \left( 1+\ctanWz \dro_0 \right)
\eea
holds.
The experimental bounds mentioned before suggest
that deviations from $\rho_0=1$ can be
treated as perturbations. In the standard approach such ``tree level''
perturbations may be included by using
\bea
\droirr \rightarrow \droirr +( 1-\rho_0^{-1})
\eea
or, in linear approximation, simply by adding
\bea
\delta \der = -\ctanWz \dro_0
\eea
where $\dro_0$ depends on the extension considered. This approach is
wrong, however. In the following we show which of the SM
contributions survive once $\rho_0$ is subject to renormalization.

Consider the low energy effective neutral current ``Fermi constant''
\bea
\wz G_{\rm NC}= \frac{\pi \al}{\mz \cose \sine }
\left(1+\delta_{\rm NC} \right)\;.
\eea
Since it is an independent parameter here and hence appears 
subtracted independently of $G_{\rm CC}=\Gmu$, no term 
$\dro$ is left over  and we have\footnote{In the notation of Ref.~\cite{AlBa}
$\dro=\epsilon_1$, $\Delta_1=\epsilon_3$ and $\Delta_2=\epsilon_2$, 
which up to normalization correspond to $T$, $S$ and $U$~\cite{stu}.} 
($s_W^2=1-c_W^2$, $c_W^2=M_W^2/M_Z^2$)
\bea
\delta _{\rm NC}
= \dal -\frac{1}{c_W^2} \Delta_1 + \delta _{\rm NC}^{\rm vertex+box}
\eea
For the leading heavy particle effects we obtain
\bea
\delta_{\rm NC}^{\rm top}&=&-K \frac{2}{3c_W^2} \ln \frac{m_t^2}{\mz} \crn
\delta_{\rm NC}^{\rm Higgs}&=&-K \frac{1}{3c_W^2} 
\left( \ln \frac{m_H^2}{\mz}-\frac{5}{3}\right)
\eea
where $K=\frac{\al}{4 \pi s_W^2}$. For the charged current amplitude we have
\bea
\wz \Gmu= \frac{\pi \al}{\mw \sine }\left(1+\delta_{\rm CC} \right)
\eea
where $\al$ and $M_W$ are renormalized as usual and $\sine$ as in the NC case.
With $\Gmu$ fixed from the $\mu$ decay rate we
have
\bea
\delta_{\rm CC}
&=& \dal -\Delta_1 +\Delta_2 + \delta _{\rm CC}^{\rm vertex+box}
\eea
The leading heavy particle effects in this case are 
\bea
\delta_{\rm CC}^{\rm top}&=&K \frac{4}{3} \ln \frac{m_t^2}{\mw} \crn
\delta_{\rm CC}^{\rm Higgs}&=&K \frac{1}{3} 
\left( \ln \frac{m_H^2}{\mw}-\frac{5}{3}\right)\;\;.
\eea
For the ratio we find
\bea
\rho=\frac{G_{\rm NC}}{\Gmu}
=\frac{\mw}{\mz \sine}\left(1-\Delta \hat{\rho}' \right)
\eea
where
$\Delta \hat{\rho}'=\delta_{\rm CC}-\delta_{\rm NC}$. 
Here the leading heavy particle terms read
\bea
\Delta \hat{\rho}^{'\mathrm{top}}&
=&K \left( \frac{4}{3}+\frac{2}{3c_W^2} \right) \ln \frac{m_t^2}{\mw} \crn
\Delta \hat{\rho}^{'\mathrm{Higgs}}&
=&-K \frac{1}{3} \frac{s_W^2}{c_W^2}
\left( \ln \frac{m_H^2}{\mw}-\frac{5}{3}\right)\;\;.
\eea
Obviously no terms proportional to $m_t^2$ (which originate in the SM
from the $W$ and $Z$ self-energies at zero momentum) have survived and
the leading heavy Higgs terms are reduced by roughly a factor 10 (!)
relative to the minimal SM. In contrast, the $m_t^2$ terms showing up
for the $Zb\bar{b}$ vertex and the observables which depend on it are
at most weakly affected due to mixing effects.

\end{document}